\documentclass[prd,aps,preprint,tightenlines]{revtex4}

\begin{document}

\newcommand{\TeV}{\,{\rm TeV}}
\newcommand{\GeV}{\,{\rm GeV}}
\newcommand{\MeV}{\,{\rm MeV}}
\newcommand{\keV}{\,{\rm keV}}
\newcommand{\eV}{\,{\rm eV}}
\def\bea{\begin{eqnarray}}
\def\eea{\end{eqnarray}}
\def\be{\begin{equation}}
\def\ee{\end{equation}}


\preprint{SNUTP 01-004, hep-ph/0101170}

\title{Lepton Flavor Violation and Radiative Neutrino Masses} 

\author{Eung Jin Chun}
\affiliation{
          Department of Physics, 
          Seoul National University,
          Seoul 151-747, Korea}


\begin{abstract}
Lepton flavor violation in various sectors of the theory
can bring important effects on neutrino masses and mixing
through wave function renormalization.  We examine general
conditions for flavor structure of radiative corrections 
producing the atmospheric and solar neutrino mass splittings
from degenerate mass patterns.  Also obtained are the mixing 
angle relations consistent with the experimental results.
\end{abstract}


\maketitle


Current data coming from the atmospheric \cite{skatm} 
and solar neutrino experiments \cite{solex} strongly indicate oscillations
among three active neutrinos following  one of the mass patterns: 
(i) hierarchical pattern with  $|m_1|, |m_2| \ll |m_3|$, 
(ii) inversely hierarchical pattern with  $|m_1|\simeq |m_2| \gg |m_3|$, 
(iii) almost degenerate pattern with   $|m_1|\simeq |m_2| \simeq |m_3|$.
In each case, one needs the mass-squared differences 
$\Delta m^2_{atm}= \Delta m^2_{32} \simeq
\Delta m^2_{31} \sim 3\times 10^{-3}$ eV$^2$  and
$\Delta m^2_{sol}=\Delta m^2_{21} \sim 10^{-4}-10^{-10}$ eV$^2$ for 
the atmospheric and solar neutrino oscillations, respectively \cite{solan}.
Here we define $\Delta m^2_{ij}\equiv m_i^2-m_j^2$ for the neutrino mass
eigenvalues $m_i$.  For the mixing angles, we take 
the standard parameterization of the neutrino mixing matrix $U$, 
\begin{equation} \label{Udef}
 U=R_{23}(\theta_1) R_{13}(\theta_2) R_{12}(\theta_3) 
 = \pmatrix { c_2c_3 & s_3c_2 & s_2 \cr
            -c_1s_3-s_1s_2c_3 & c_1c_3-s_1s_2s_3 & s_1c_2 \cr
         s_1s_3-c_1s_2c_3 & -s_1c_3-c_1s_2s_3 & c_1c_2 \cr },
\end{equation}
where $R_{ij}(\theta_k)$ is the rotation 
in the $ij$ plane by the angle $0\leq \theta_k \leq \pi$ and $c_k =
\cos\theta_k$, $s_k=\sin\theta_k$. In our discussion, 
we neglect CP-violating phases. 
The atmospheric neutrino data require the $\nu_\mu-\nu_\tau$
oscillation amplitude 
$A_{atm}= c_2^2\sin^22\theta_1 \approx 1$ implying nearly
maximal mixing $\theta_3\approx \pi/4$.  
For the solar neutrino oscillation, the $\nu_e-\nu_{\mu,\tau}$ oscillation
amplitude $A_{sol}=c_2^2\sin^22\theta_3$ can take either 
large $\sim 1$ or small $\sim 10^{-3}$ value (that is, $\theta_1\sim \pi/4$ 
or $\theta_1\ll 1$) depending on the solutions to the solar neutrino problem.
The mixing element $U_{e3}$ is constrained by the reactor experiment 
on the $\bar{\nu}_e$ disappearance \cite{chooz}; 
$|U_{e3}|=|s_2| \lesssim 0.2$ for $\Delta m^2_{32}$ 
allowed by the atmospheric neutrino data.
There is also a limit on the neutrino mass element itself
coming from the neutrinoless double beta decay experiments, 
$|M_{ee}|=|\sum_i m_i U_{ei}^2|< 0.2$ eV \cite{0nbb}.  
This bound is particularly important for the degenerate pattern (iii) 
where $m_i^2 \gg \delta m^2_{atm}$ is possible.
Note that we restricted ourselves to oscillations among three active
neutrinos disregarding the LSND results \cite{lsnd} waiting for a
confirmation in the near future.

Understanding the origin of neutrino masses and mixing is one of
fundamental problems in physics beyond the standard model.
It concerns with the leptonic flavor structure of the theory.
As we know, small Majorana neutrino mass textures 
(at tree level) would be attributed to the effective higher 
dimensional operator $L_\alpha  L_\beta  H_2 H_2$ where $L_\alpha$ 
denotes the lepton doublet with the flavor index $\alpha=e,\mu,\tau$ 
and $H_2$ is the Higgs field coupling to up-type quarks.
This operator breaks the total lepton number by 2 units ($\Delta L=2$).
There can be other sectors breaking individual lepton number 
(with $\Delta L=0$) which give important contributions to
the  neutrino mass matrix through radiative corrections.  
The best-known example is the renormalization group effect of 
the charged lepton Yukawa couplings \cite{rge1}, 
which has been studied extensively in recent years \cite{rge2}.  
Such an effect provides potentially important origin of tiny neutrino mass 
splittings, of course, depending on  specific tree level neutrino 
mass textures.

In this regard, 
it is an interesting question whether the desired mass splittings
for the degenerate mass patterns  (ii) and (iii)
can arise from radiative corrections bringing the effect
of lepton flavor violation in various sectors of the theory.
The purpose of this work is to investigate 
general conditions for radiative corrections to produce the 
observed neutrino mass splittings without resorting to any specific
models for flavor structure beyond the Yukawa sector. 
Assuming that various radiative correction contributions are 
hierarchical, we will identify appropriate flavor structure of
the dominant or subdominant contribution which are able to generate
the atmospheric or solar neutrino mass splitting.
For these loop corrections, we also derive mixing angle relations
consistent with the experimental data.  
Our results would be useful for constructing models 
of degenerate neutrinos.

\medskip

The general form of the loop-corrected neutrino mass matrix 
due to wave function renormalization is given by
\begin{equation} \label{renfl}
 M_{\alpha \beta} = m_{\alpha\beta} +{1\over2} \left(
   m_{\alpha\gamma} I_{\gamma\beta}
  +I_{\alpha\gamma} m_{\gamma\beta}  \right)\,,
\end{equation}
where $m_{\alpha\beta}$ is the tree level neutrino mass matrix and 
$I_{\alpha\beta}$ is the loop contribution.
Note that the above formula is written in the flavor basis 
where charged lepton masses are diagonal.  
Often, it is convenient to re-express 
Eq.~(\ref{renfl}) in the tree level mass basis of neutrinos;
\begin{equation} \label{renms}
 M_{ij}=m_i \delta_{ij}+{1\over2} (m_i+m_j) I_{ij} \,,
\end{equation}
where $m_i$ is the tree level mass eigenvalue and  
the loop correction factor $I_{ij}$ 
is related to $I_{\alpha\beta}$ by the equation,
$I_{ij} = \sum_{\alpha\beta} I_{\alpha\beta} U_{\alpha i} U_{\beta j}$,
where $U$ is the tree level diagonalization matrix parametrized as in
Eq.~(\ref{Udef}).  In this paper, we have nothing to mention about
the origin of the flavor structure of $m_{\alpha\beta}$.
Given $m_{\alpha\beta}$ yielding 
the degenerate mass pattern of the type (ii) or (iii), we will examine the
flavor structure of the loop factor $I_{\alpha\beta}$ which can
produce the atmospheric and/or solar neutrino masses and mixing.
Before coming to our main point, it is instructive to see where
the loop correction $I_{\alpha\beta}$ can come from.

\medskip

To illustrate how the flavor structure of the loop correction
$I_{\alpha\beta}$ can arise,
we consider one of the most popular model beyond the standard model, namely,
the supersymmetric standard model.
Like the standard model, it has an inevitable flavor violation in the 
Yukawa sector with the superpotential,
\begin{equation} \label{Wyuk}
 W \ni h_\alpha H_1 L_\alpha E^c_{\alpha} \,,
\end{equation}
where $H_1$, $L_\alpha$ and $E^c_\alpha$ denote the Higgs, lepton doublet
and singlet superfields.  At the leading log approximation,  
the Yukawa terms in Eq.~(\ref{Wyuk}) give rise to \cite{rge1,rge2}
\begin{equation} \label{IYuk}
 I_{\alpha\alpha} \approx -{h_\alpha^2 \over 8\pi^2} 
            \ln\left(M_X\over M_Z\right) \,,
\end{equation}
where $M_X$ denote a fundamental scale generating the above-mentioned
effective operator and the Z boson mass $M_Z$ represents the weak scale.

The soft supersymmetry breaking terms included in this model
are also potentially important sources of flavor violation.
Those terms include sfermion masses and trilinear A-terms which
are generically nonuniversal and flavor dependent;  
\begin{equation}\label{softs}
 V_{soft} \ni m^2_{\alpha\beta^*} L_\alpha L_\beta^\dagger 
 + A_{\alpha\beta} H_1 L_\alpha E^c_\beta +   h.c. \,,
\end{equation}
where we use the same notation for the scalar components of superfields.
The effect of slepton masses has been first discussed in Ref.~\cite{hoza}.
The off-diagonality and non-degeneracy in diagonal masses of 
$\tilde{m}^2_{\alpha\beta^*}$ give rise to
\begin{equation}
I_{\alpha\beta} = {g^2 \over 8\pi^2} \delta^l_{\alpha\beta} 
 f(\tilde{m}^2_{\alpha\alpha}; \tilde{m}^2_{\beta\beta}) \,,
\end{equation}
where $\delta^l_{\alpha\beta}= 
\tilde{m}^2_{\alpha\beta}/(\tilde{m}_{\alpha\alpha}\tilde{m}_{\beta\beta})$
for $\alpha\neq\beta$ (assuming $\delta^l_{\alpha\beta}\ll1$),
$\delta^l_{\alpha\alpha}=1$ and $f$ is an appropriate loop function
of order one.
In the similar way, the effect of the A-terms 
comes from one-loop diagrams with wino/zino and slepton exchange
generating
\begin{equation}
 I_{\alpha\beta} \approx 
 {g^2\over 8\pi^2} {A_{\alpha\gamma} A^*_{\beta\gamma} 
 \langle H_1\rangle^2 \over \tilde{m}^4} \,,
\end{equation}
where $\tilde{m}$ is a typical mass of the sparticles 
running inside the loop.  
In general, the A-terms are not proportional 
to the (charged lepton) Yukawa couplings,
\newcommand{\npropto}{\propto\hspace{-2ex}/\hspace{1ex}}
$A_{\alpha\beta} \npropto h_{\alpha\beta}=\delta_{\alpha\beta}h_\alpha$,  
This can lead to important new contributions to lepton flavor 
changing loop corrections as above.  
However, we expect in generic models that 
$A_{\alpha\beta} \lesssim \tilde{m} h_\tau$ 
and thus the sizes of their loop effects are at most 
comparable to those of the tau Yukawa coupling. 

Lepton flavor violation can also appear in the superpotential through
R-parity and lepton number violating trilinear terms;
\begin{equation} \label{RW}
 W \ni \lambda_{\alpha\beta\gamma} L_\alpha L_\beta E^c_\gamma
     + \lambda'_{\alpha\beta\gamma} L_\alpha Q_\beta D^c_\gamma  \,,
\end{equation}
where $Q, D^c$ are doublet and down-type singlet quark superfields.
The R-parity violating couplings can participate in
the renormalization group equation like the charge lepton Yukawa couplings,
and thus  we get for the couplings $\lambda$,
\begin{equation}
 I_{\alpha\beta} \approx 
 -{ \lambda_{\alpha\gamma\delta}\lambda^*_{\beta\gamma\delta}
 \over 8\pi^2} \ln({M_X\over M_Z})\,.
\end{equation}
For certain combinations of R-parity violating couplings \cite{Rbounds}, 
the current experimental bounds are weak enough to give a sizable 
loop correction $I_{\alpha\beta}$.   In particular, if one takes only one
dominant coupling $\lambda_{\alpha\gamma\delta}$ which can be as large as 
order one,   one can have a very large loop correction $I_{\alpha\alpha}$. 
With the generic R-parity violation (\ref{RW}), of course, 
one may have a finite loop correction to neutrino 
masses like $\delta m_{\alpha\beta} \propto \lambda_{\alpha\gamma\delta} 
\lambda_{\beta\delta\gamma} h^e_{\gamma}h^e_\delta$  \cite{Rmass}
which is not a topic of the present investigation.  

\medskip

As we have seen above, there could be rich sources for sizable 
loop corrections $I_{\alpha\beta}$ in a general class of models.  
In the below, we discuss the effect of
general flavor-dependent loop contributions $I_{\alpha\beta}$  for various
degenerate mass patterns at tree level.  
Let us start with analyzing the conditions to produce 
the desired mass splittings for the atmospheric {\it as well as} 
solar neutrino oscillations in the case of 
the fully degenerate mass patterns; $|m_1|=|m_2|=|m_3|$.  
For this,  we look first for the possible loop corrections 
giving rise to the atmospheric mass splitting $\Delta m^2_{32}/2m_0^2$ at
the leading order correction.  In our analysis, it is assumed that
loop corrections $I_{\alpha\beta}$  take hierarchical values and thus
the leading order splitting is dominated by one specific $I_{\alpha\beta}$.
For the fully degenerate pattern, there are the following possibilities
depending on the CP-conserving phases of the tree level mass eigenvalues;
$$ {\rm I.} \quad 
 (m_1,m_2,m_3)=m_0(-1,-1,1)\quad{\rm or}\quad m_0(\pm1, \mp1,1)$$ 
which we call the 1-2 or 1-3 (2-3) degeneracy, respectively.
The loop correction can induce nonvanishing off-diagonal components 
$M_{ij}$ as far as $m_i+m_j\neq0$ (\ref{renms}).  Then,
as discussed in Ref.~\cite{hoza}, one has a freedom to define the tree level 
mixing angles in the matrix $U$ in such a way that 
$I_{ij}=I_{\alpha\beta}U_{\alpha i} U_{\beta j} =0$ 
due to the exact degeneracy $m_i=m_j$.  It is a simple manner to show this
explicitly.  In the $i$-$j$ plane  with tree-level degeneracy, 
the radiatively corrected mass matrix is diagonalized by the rotation
$R_{ij}(\phi)$ where $\phi$ is given by
\begin{equation} \label{rephi}
 \tan2\phi = {2I_{ij} \over I_{jj} - I_{ii}} \,.
\end{equation}
On the other hand, we are free to choose our tree level mixing matrix:
$U \to U'=UR_{ij}(\phi)$ where $R_{ij}(\phi)$ is given by
$R_{ij,kl}(\phi)=c_\phi(\delta_{ik}\delta_{il}+\delta_{jk}\delta_{jl})
+s_\phi(\delta_{ik}\delta_{jl}-\delta_{jk}\delta_{il})$.  Then 
the loop factor $I'_{ij} = I_{\alpha\beta}U'_{\alpha i} U'_{\beta j}$ written
in terms of the new mixing matrix $U'$ becomes
$I'_{ij}=[2I_{ij} \cos2\phi+(I_{ii}-I_{jj})\sin2\phi]/2$ which vanishes
due to the relation (\ref{rephi}). Note that this conclusion holds for
arbitrary sum of $I_{\alpha\beta}$.

With this {\it properly} defined mixing matrix $U$ satisfying
the {\it  mixing angle relation} $I_{ij}=0$, the nonvanishing
mass-squared differences generated by loop correction can be written as
\begin{eqnarray} \label{mscor}
 \Delta m^2_{21} &=& 2m_0^2(I_{22}-I_{11}) = 
  2m_0^2 I_{\alpha\beta}(U_{\alpha2}U_{\beta2}-U_{\alpha1}U_{\beta1} ) 
           \nonumber\\
 \Delta m^2_{32} &=& 2m_0^2(I_{33}-I_{22}) =
  2m_0^2 I_{\alpha\beta}(U_{\alpha3}U_{\beta3}-U_{\alpha2}U_{\beta2} ) \,.
\end{eqnarray}
When a specific component $I_{\alpha\beta}$ gives the  dominant 
contribution in Eq.~(\ref{mscor}), 
we need to have $(U_{\alpha2}U_{\beta2}-U_{\alpha1}U_{\beta1} )/
(U_{\alpha3}U_{\beta3}-U_{\alpha2}U_{\beta2} )  \lesssim 
\Delta m^2_{sol}/\Delta m^2_{atm} $, that is, 
$U_{\alpha2}U_{\beta2} \approx U_{\alpha1}U_{\beta1}$ is required 
to get $ |m^2_{32}| \gg   |m^2_{21}|$.   Based on these properties, 
we are now ready to consider the effect of a loop correction 
$I_{\alpha\beta}$.

(a) $I_{\alpha\alpha}$ dominance:
For the $i$-$j$ degeneracy, the relation 
$I_{ij} \propto U_{\alpha i} U_{\alpha j}=0$ implies
$U_{\alpha i}=0$ or $U_{\alpha j}=0$.
As an immediate consequence, we find that the $1$-$2$ 
degeneracy cannot work for any $\alpha$ since it gives 
$U^2_{\alpha1}  \simeq U^2_{\alpha2}\simeq 0$ and 
thus $U^2_{\alpha3}\simeq 1$ contradicting  with the empirical results,
$U^2_{e3}\ll1$ and $U^2_{\mu3}\simeq U^2_{\tau3}\simeq 1/2$.
In fact, the only possible mixing angle relation is $U_{e3}=s_2=0$, 
which can be realized only with the $I_{ee}$ dominance combined 
with the 1-3 or 2-3 degeneracy.  In this case, we get from Eq.~(\ref{mscor})
\begin{equation} \label{msee}
{\Delta m^2_{21} \over \Delta m^2_{32}} \simeq {\cos2\theta_3\over s_3^2}
\end{equation}
which requires the maximal mixing of the solar neutrinos,
$\cos2\theta_3\ll1$.

(b) $I_{\alpha\beta}$ dominance ($\alpha\neq\beta$): 
As discussed below Eq.~(\ref{mscor}),
it is useful to notice that 
we need  $0\neq U_{\alpha2}U_{\beta2} \simeq 
U_{\alpha1}U_{\beta1} \simeq -U_{\alpha3}U_{\beta3}/2$ (the last relation
comes from the orthogonality condition
$\sum_i U_{\alpha i}U_{\beta i}=\delta_{\alpha\beta}$) to yield   
$\Delta m^2_{21}/\Delta m^2_{32} \approx 2(I_{22}-I_{11})/3I_{22}$.
Then, we can easily rule out the case  $(\alpha\beta)=(e\mu)$ or $(e\tau)$
from the simple observation that 
$$
{\Delta m^2_{21} \over \Delta m^2_{32}} \simeq
    {2\over 3}\left[ \cos2\theta_3+ {c_1\over s_1 s_2} \sin2\theta_3\right]
  \quad{\rm or}\quad  {2\over 3}\left[ \cos2\theta_3- {s_1\over c_1 s_2} 
   \sin2\theta_3\right]
$$
which cannot be made small enough for realistic values of the mixing 
angles satisfying $\theta_1 \simeq \pi/4$, $\theta_2 \ll 1$ and 
$\theta_3 \simeq \pi/4$ or $\theta_3 \ll 1$.  
On the other hand, for $(\alpha\beta)=(\mu\tau)$, we get 
\begin{equation} \label{msmt}
{\Delta m^2_{21} \over \Delta m^2_{32}} \simeq
  -{2\over3c_2^2}\left[ (1+s_2^2)\cos2\theta_3
  + 2 s_2 \sin{2\theta_3} \cot{2\theta_1} \right]
\end{equation}
which becomes vanishingly small for $s_2 \cos2\theta_1  \ll 1$ 
and $\cos2\theta_3 \ll 1$. 
Now we have to check if the mixing angle relation fixed by the condition
$I_{ij}=0$ for certain combination of  $(ij)$ can be consistent with our  
consideration.  As shown in Ref.~\cite{hoza},  the 1-3 or 2-3 degeneracy
gives again the desired relation respectively,
\begin{equation} \label{s2mt}
s_2 = - \cot2\theta_1 \tan\theta_3\quad{\rm or} \quad
 s_2=   \cot2\theta_1 \cot\theta_3 \,.
\end{equation}
which relates the smallness of the angle $\theta_2$ with
the large atmospheric neutrino mixing.

\begin{table}
\begin{center}
\begin{tabular}{|c|c|c|c|} \hline
$r$ & $\Delta m^2_{32}/m_0^2 $ & 
$\Delta m^2_{21}/m_0^2 $ & $s_2$  \\\hline
$I_{ee}/I_{\mu\tau}$ 
&  & 0 & 0 \\
$I_{\mu\mu}/I_{\mu\tau}$ 
&  & $I_{\mu\tau} r^2 \sin2\theta_1/2$ & $-rs_3/2c_3$ \\
$I_{\tau\tau}/I_{\mu\tau}$   & $I_{\mu\tau} \sin2\theta_1$
  & $I_{\mu\tau}  r^2 \sin2\theta_1/2$ & $rs_3/2c_3$ \\
$I_{e\mu}/I_{\mu\tau}$ 
&  & $I_{\mu\tau}  r c_1 c_2 \sin2\theta_3$ & $r/2c_1c_2$ \\
$I_{e\tau}/I_{\mu\tau}$ 
&  & $-I_{\mu\tau}  r s_1 c_2 \sin2\theta_3$ & $r/2s_1c_2$ \\
\hline
$I_{\mu\mu}/I_{ee}$ 
&  & $-I_{ee} r^2/2$ & $rc_1s_1s_3/c_3$ \\
$I_{\tau\tau}/I_{ee}$   
& & $-I_{ee} r^2/2$ & $-rc_1s_1s_3/c_3$ \\
$I_{e\mu}/I_{ee}$ & $-I_{ee}$
  & $I_{ee}  r c_1 c_2 \sin2\theta_3$ & $-rs_1/c_2$ \\
$I_{e\tau}/I_{ee}$ 
&  & $-I_{ee}  r s_1 c_2 \sin2\theta_3$ & $-rc_1/c_2$ \\
$I_{\mu\tau}/I_{ee}$ 
&  &  0 & 0 \\
\hline
\end{tabular}
\end{center}
\caption{Possibilities of radiative generation of
$\Delta m^2_{atm}$ and $\Delta m^2_{sol}$ in the case of the
degeneracy, $m_1=-m_2=m_3$.
The upper and lower box correspond to the cases of 
the $I_{\mu\tau}$ and $I_{ee}$ dominance, respectively.
The last column shows the correction to $U_{e3}=s_2$ from exact bimaximal
mixing given subdominant $I_{\alpha\beta} \ll I_{\mu\tau}$ or $I_{ee}$.}  
\end{table}

In sum, the small mass splitting for the atmospheric neutrinos
can be obtained {\it only} with the dominant loop correction of $I_{ee}$ 
or $I_{\mu\tau}$ in the case of the 1-3 or 2-3 degeneracy.
Furthermore, this picture can be consistent only with {\it bimaximal} 
mixing $\theta_1, \theta_3 \simeq \pi/4$ and $s_2 \ll 1$ fixed by
the mixing angle relation $I_{13}$ or $I_{23}=0$.  
Note that all of these are fairly consistent with the neutrinoless
double beta decay bound as we have 
$|M_{ee}|=|m_0(c_2^2\cos2\theta_3\pm s_2^2)| \ll |m_0|$.

Let us turn to the solar neutrino mass splitting.  As can be seen from
Eqs.~(\ref{msee}) and (\ref{msmt}), the right values for $\Delta m^2_{21}$
may arise if the mixing anlges satisfy
 $\cos2\theta_3 \sim \cos^22\theta_1$ $\sim 
\Delta m^2_{sol}/\Delta m^2_{atm}$ 
where the second relation is applied only to the $I_{\mu\tau}$ dominance.
However, it appears unnatural to arrange such  small values for the
tree level mixing angles.  It would be more plausible to imagine 
the situation of exact bimaximal mixing ($\cos2\theta_1=\cos2\theta_3 =0$) 
imposed by certain symmetry in tree level mass matrix.  Then, the 
solar neutrino mass splitting could be generated by a smaller loop 
correction $I_{\alpha\beta}$ other than $I_{ee}$ or $I_{\mu\tau}$.
Including now in Eq.~(\ref{mscor}) this subdominant contribution,
one can find the deviations from the leading results,
$\Delta m^2_{21}=0$ and  $s_2=0$.
The result of our calculation is summarized in Table I
in the case of the 1-3 degeneracy. We have shown
the explicit angle dependences to notify the sign of mass-squared difference
which might be distinguishable by the solar neutrino MSW effect.
Similar result can be obtained for the 2-3 degeneracy.
A few remarks are  in order. The desired size of the loop correction 
for the atmospheric neutrino mass-squared difference
$I_{ee,\mu\tau} \approx \Delta m^2_{atm}/m_0^2$.
Thus, the degenerate mass $m_0\sim 1$ eV of cosmological interests needs 
$I_{ee, \mu\tau} \sim 10^{-3}$
which  is a reasonable value for radiative corrections.
As can be seen in Table I,  the ratio $\Delta m^2_{sol}/\Delta m^2_{atm}$ 
is roughly given by $r$ or $r^2$ depending on the flavor structure of 
the radiative corrections whereas $s_2\sim r$ for any cases.  
If the large angle MSW solution to the solar neutrino problem 
is realized, one needs $r$ or $r^2$ of the order $10^{-2}$, that is,
a loop correction $I_{\alpha\beta}$ should be smaller than 
$I_{ee, \mu\tau}$ by factor of $10^{-2}$ or $10^{-1}$.  
For the latter case, we get $s_2 \sim 0.1$ which is within 
the reach of future experiments.
We note that a supersymmetric model realizing the case with the 
$I_{ee}$ dominance and $r=I_{\tau\tau}/I_{ee}$ has been worked out
in Ref.~\cite{valle}. 

\medskip

Another possibility is to have the atmospheric neutrino
mass splitting given at tree level and the smaller splitting for
the solar neutrino mass is driven by loop corrections.  This includes
almost full degeneracy $|m_1|=|m_2|\simeq |m_3|$
and inverse hierarchy $|m_1|=|m_2|\gg |m_3|$, both of which can be 
parametrized as
$${\rm II.}\quad (m_1,m_2,m_3)=m_0(1,\pm1,z)\,,$$
where $z=\pm1+\delta_a$ with $|\delta_a| = \Delta m^2_{atm}/2m_0^2$
for the almost full degeneracy, or  $|z|\ll 1$ with 
$m_0^2=\Delta m^2_{atm}$ for the inverse hierarchy.
We consider the two cases, $m_1=\pm m_2$, separately.

\begin{table}
\begin{center}
\begin{tabular}{|c|c|c|} \hline
$I_{\alpha\beta}$ & $I_{12}=0$ & $(I_{22}-I_{11})/I_{\alpha\beta}$  \\ \hline
$I_{ee}$ &  $c_2^2\sin2\theta_3=0$ & $-c_2^2\cos2\theta_3$  \\
$I_{\mu\mu}$ & 
 ${s_2 \over c_1^2+s_1^2s_2^2} = \pm{\sin2\theta_3 \over \sin2\theta_1}$ &
 $c_1^2\cos2\theta_3-s_2\sin2\theta_1\sin2\theta_3$  \\
$I_{\tau\tau}$ & 
 ${s_2 \over s_1^2+c_1^2s_2^2} = \pm{\sin2\theta_3 \over \sin2\theta_1}$ &
 $s_1^2\cos2\theta_3+s_2\sin2\theta_1\sin2\theta_3$  \\
$I_{e\mu}$ & 
 $s_2=\cot\theta_1\cot2\theta_3$ &
 $c_2(+c_1\sin2\theta_3+s_1s_2\cos2\theta_3)$  \\
$I_{e\tau}$ & 
 $s_2=-\tan\theta_1\cot2\theta_3$ &
 $c_2(-s_1\sin2\theta_3 + c_1s_2\cos2\theta_3)$  \\
$I_{\mu\tau}$ &
 ${2s_2 \over 1+s_2^2} = \tan2\theta_1\tan2\theta_3$  &
 $-{1\over2}\sin2\theta_1\cos2\theta_3-s_2\cos2\theta_1\sin2\theta_3$ 
 \\ \hline
\end{tabular}
\end{center}
\caption{The mixing angle relation  and the loop contribution
to $\Delta m^2_{21}$  for each dominant $I_{\alpha\beta}$
in the case of the degeneracy, $m_1=m_2 \neq m_3$.  }
\end{table}

(a) $m_1=m_2=m_0$:
As discussed before, the mixing angles satisfy $I_{12}=0$ and the
leading contribution to  $\Delta m^2_{21}$ is given by
$2m_0^2(I_{22}-I_{11})$.  These two quantities are presented in 
Table II.  One can realize that the $I_{ee}$ dominance does not work at all.
For $(\alpha\beta)=(\mu\mu), (\tau\tau), (\mu\tau)$, only the {\it small}
solar mixing angle is consistent since the
mixing angle relation $s_2\propto \sin2\theta_3$ has to be put.
On the contrary, for $(\alpha\beta)=(e\mu), (e\tau)$, the {\it large}
solar mixing can only be allowed since $s_2 \propto \cos2\theta_3$. 
Imposing these mixing angle relations, one can see that 
$I_{22}-I_{11}$ in Table II does not vanish for any $(\alpha,\beta)$,
and thus $\Delta m^2_{sol} \approx m_0^2 I_{\alpha\beta}$.
In the case of $|z|\ll 1$, we thus need $I_{\alpha\beta} \approx \Delta 
m^2_{sol}/\Delta m^2_{atm} \lesssim 10^{-2}$ where the approximate equality
is for the large angle MSW solution and may be a little large value for 
a radiative correction.
Here it is worth noting that {\it e.g.}, for the $I_{\tau\tau}$ dominance,
we have $\sin^22\theta_3 \ll 1$  and 
\begin{equation} \label{sgn}
\Delta m^2_{21} \simeq 2m_0^2 I_{\tau\tau} s_1^2 (c_3^2-s_3^2) \,.
\end{equation}
For the small solar neutrino  mixing to work, 
we need $s_3^2 \ll 1$ for $\Delta m^2_{21}>0$ 
(or  $\Delta m^2_{21}<0$ for $c_3^2\ll1$), which requires 
from Eq.~(\ref{sgn}) that $I_{\tau\tau} >0$.   Therefore, $I_{\tau\tau}$
given in Eq.~(\ref{IYuk}) for the supersymmetric standard model does 
{\it not} fulfill this condition whereas the usual standard model with 
$I_{\tau\tau} \approx h_\tau^2 \ln(M_X/M_Z)/16\pi^2$ can work.

(b) $m_1=-m_2=m_0$:
In this case, no mixing angle relation is imposed.
Including the effect of the off-diagonal elements 
$M_{13}$ and $M_{23}$ generated from loop correction, we find
\begin{equation} \label{extra}
 \Delta m^2_{21}=2m_0^2\left[I_{22}-I_{11}
      +{1\over2}{(z-1)^2 \over (z+1)} I^2_{23}
      +{1\over2}{(z+1)^2 \over (z-1)} I^2_{13}\right] \,,
\end{equation}
where $I_{22}-I_{11}$ is given in Table II for each $I_{\alpha\beta}$.
Depending on the mixing angles fixed at tree level, the
leading contribution to $\Delta m^2_{sol}$ may come from $I_{22}-I_{11}$
or the next terms in Eq.~(\ref{extra}).  That is, we need to have
$\Delta m^2_{sol}/\Delta m^2_{atm}\sim I_{\alpha\beta}/\delta_a$ 
or $I_{\alpha\beta}^2/\delta_a^2$ for $z=\pm1+\delta_a$, and 
$\Delta m^2_{sol}/\Delta m^2_{atm}\sim I_{\alpha\beta}$ or 
$I_{\alpha\beta}^2$ for  $|z|\ll1$.   Therefore, the values of
$\Delta m^2_{sol}$ for various solutions to solar neutrino problem can be 
obtained with appropriate values of $I_{\alpha\beta}$ and $\delta_a$.
From Table II, one can see
that the leading term $I_{22}-I_{11} \propto I_{\alpha\beta}$ vanishes
for the exact bimaximal mixing with $s_2=0$ and $\cos2\theta_3=0$ in the case
of $(\alpha\beta)=(ee),(\mu\mu), (\tau\tau)$ and $(\mu\tau)$.  
Furthermore, for $(\alpha\beta)=(ee)$ and $(\mu\tau)$, $I_{13}$ and
$I_{23}$ also vanish and thus no splitting can arise at one-loop level.
We would like to stress that either the large or small mixing solution to 
the solar neutrino problem can be realized as we have no 
mixing angle relation imposed.  For the small mixing solution,
the solar neutrino mass-squared difference has further suppression
by small mixing angles as $I_{22}-I_{11} \sim I_{\alpha\beta}\sin2\theta_3$
or $I_{\alpha\beta}s_2$ in the case of
$(\alpha\beta)=(e\mu)$ or $(e\tau)$.

\medskip

In conclusion, we have discussed the general conditions for generating
small neutrino mass splittings from the effect of one-loop wave 
function renormalization of degenerate neutrino masses at tree level.
Without assuming any specific model on the structure of lepton
flavor violation other than the tree level neutrino mass sector,
we identified the flavor dependences of the loop factors
which can give rise to the neutrino mass-squared difference and mixings
required by the atmospheric and solar neutrino data.
For the fully degenerate pattern $|m_1|=|m_2|=|m_3|$, the atmospheric
neutrino mass splitting can be obtained when the dominant loop correction
comes from $I_{ee}$ or $I_{\mu\tau}$  in the cases of $m_1=m_3$ and $m_2=m_3$.
In each case, the smaller loop corrections are examined  to generate
the solar neutrino mass splitting.  For all the cases,  it turns out that 
the so-called bimaximal mixing can only be realized.
For the partially degenerate case $|m_1|=|m_2|\neq |m_3|$, 
we identified $I_{\alpha\beta}$ with which 
the desired mass splitting and the (large or small) mixing angle
for the solar neutrino oscillations can be obtained.
Our results may be useful for explicit model building along this line.

\medskip

{\bf  Acknowledgement}:
The author is supported by the BK21 program of Ministry of Education.


\begin{thebibliography}{99}
\def\plb#1#2#3{Phys.\ Lett.\       {\bf B#1}  (#3) #2}
\def\npb#1#2#3{Nucl.\ Phys.\       {\bf B#1}  (#3) #2}
\def\prd#1#2#3{Phys.\ Rev.\        {\bf D#1}  (#3) #2}
\def\prl#1#2#3{Phys.\ Rev.\ Lett.\ {\bf #1}   (#3) #2}
\def\mpl#1#2#3{Mod.\ Phys.\ Lett.\ {\bf A#1}  (#3) #2}
\def\rep#1#2#3{Phys.\ Rep.\        {\bf #1}   (#3) #2}
\def\sci#1#2#3{Science             {\bf #1}   (#3) #2}
\def\astro#1#2#3{Astrophys.\ J.\   {\bf #1}   (#3) #2}
\def\epj#1#2#3{Eur.\ Phys.\ J.\   {\bf C#1} (#3) #2}
\def\jhep#1#2#3{JHEP {\bf #1} (#3) #2}
\def\ptp#1#2#3{Prog.\ Theor.\ Phys.\ {\bf #1} (#3) #2}

\bibitem{skatm}  Super-Kamiokande Collaboration, Y. Fukuda {\it et al.},
             \prl{81}{1562}{1998}.
\bibitem{solex} Homestake Collaboration, B.T. Cleveland {\it et al.},
                 \astro{496}{505}{1998};
                 Kamiokande Collaboration, K.S. Hirata {\it et al.},
                 \prl{77}{1683}{1996};
                 GALLEX Collaboration, W. Hampel {\it et al.},
                 \plb{388}{384}{1996}; 
                 SAGE Collaboration, D.N. Abdurashitov {\it et al.},
                 \prl{77}{4708}{1996};
                 Super-Kamiokande Collaboration, Y. Fukuda {\it et al.},
                 \prl{82}{2430}{1999}.
\bibitem{solan} 
For a recent global analysis, see M.C. Gonzalez-Garcia {\it et al.},
hep-ph/0009350.
\bibitem{chooz} CHOOZ Collaboration, A. Apollonio {\it et al.},
                \plb{420}{397}{1998}; Palo Verde experiment, 
                \prl{84}{309}{2000}.
\bibitem{0nbb} 
      L. Baudis {\it et al.}, hep-ex/9902014.
\bibitem{lsnd}  LSND Collaboration, C. Athanassopoulos, {\it et.al.}, 
      Phys. Rev. Lett. {\bf 75} (1995) 2650; {\bf 81} (1998) 1774.

\bibitem{rge1} 
    P.H. Chankowski and Z. Pluciennik, \plb{316}{312}{1993};
    K. Babu, C. N. Leung and J. Pantaleone, \plb{319}{191}{1993};
    M. Tanimoto, \plb{360}{41}{1995}.
\bibitem{rge2}
    J. Ellis and S. Lola, \plb{458}{310}{1999}; 
    N. Haba {\it et al.}, \ptp{103}{367}{2000}; \epj{10}{177}{1999};
     \epj{14}{347}{2000};
    J. A. Casas, J. R.  Espinosa, A. Ibarra and I. Navarro, 
    \npb{556}{3}{1999}; \npb{569}{82}{2000}; \jhep{9909}{015}{1999};
    \npb{573}{652}{2000};
    E. Ma, J. Phys. {\bf G25}, 97 (1999);
    R. Barbieri, G.G. Ross and A. Strumia, JHEP 9910:020 (1999);
    P.H. Chankowski, W. Krolikowski and S. Pokorski, \plb{473}{109}{2000};  
    A.S. Dighe and A.S. Joshipura, hep-ph/0010079.

\bibitem{Rbounds} G. Bhattacharyya, hep-ph/9709395;
 B.C. Allanach, A. Dedes and H.K. Dreiner, \prd{60}{075014}{1999}.

\bibitem{Rmass}
 For recent works, see {\it e.g.}, E.J. Chun and S.K. Kang, 
 \prd{61}{075012}{2000}; S. Davidson and M. Losada, hep-ph/0010325.

\bibitem{hoza}
    E.J. Chun and S. Pokorski, \prd{62}{053001}{2000}.

\bibitem{valle} P.H. Chankowski, A. Ioannisian, S. Pokorski and
  J.W.F.  Valle, hep-ph/0011150.


\end{thebibliography}
\end{document}